# Role of magnetic skyrmions for the solution of the shortest path problem

Riccardo Tomasello[1*], Anna Giordano[2,3], Francesca Garescì[4], Giulio Siracusano[2], Salvatore De Caro[4], Caterina Ciminelli[5], Mario Carpentieri[5], Giovanni Finocchio[2,3*]

[1]*Institute of Applied and Computational Mathematics, FORTH, GR-70013, Heraklion-Crete, Greece*

[2]*Department of Mathematical and Computer Sciences, Physical Sciences and Earth Sciences, University of Messina, I-98166 Messina, Italy*

[3]*Istituto Nazionale di Geofisica e Vulcanologia, Via di Vigna Murata 605, I-00143 Roma, Italy*

[4]*Department of Engineering, University of Messina, Messina, Italy*

[5]*Department of Electrical and Information Engineering, Politecnico di Bari, Bari, Italy*

**Abstract**

Magnetic skyrmions are emerging as key elements of unconventional operations having unique properties such as small size and low current manipulation. In particular, it is possible to design skyrmion based neurons and synapses for neuromorphic computing in devices where skyrmions move along the current direction (zero skyrmion Hall angle). Here, we show that, for a given graph, skyrmions can be used in optimization problems facing the calculation of the shortest path. Our tests show a solution with the same path length as computed with the Dijkstra's Algorithm. In addition, we also discuss how skyrmions act as positive feedback on this type of problems giving rise to a self-reinforcement of the path which is a possible solution.

Corresponding authors: rtomasello@iacm.forth.gr, gfinocchio@unime.it

# 1. Introduction

Magnetic skyrmions are particle-like magnetization textures[1,2] characterized by a non-trivial topology and usually stabilized in presence of a sufficiently large Dzyaloshinskii–Moriya interaction (DMI)[3,4]. Skyrmion solutions were initially obtained by Skyrme[5] in the non-linear field theory, while recently they have been achieved in magnetic systems where Néel-type[6–8] and Bloch-type skyrmions[9–12] can be observed. The increasing interest on skyrmions is mainly due to the fact that they can be manipulated by electrical currents, which opens the way to a number of potential applications[1,2,13]. In particular, the skyrmion motion can be driven by either the spin-transfer torque (STT)[6,7,12,14] or the spin-Hall effect (SHE)[6,7,15]. A major feature of the current-driven skyrmion motion is that its trajectory does not coincide with the direction of the applied electrical current, but it is characterized by an in-plane deflection angle, i.e. the skyrmion Hall angle (SHA)[16,17]. The control of the SHA is crucial for all the skyrmions applications based on their in-plane motion, where skyrmions would be unavoidably driven towards the sample edges and consequently annihilated. The suppression of the SHA is achievable by balancing the Magnus force in two coupled skyrmions with opposite topological charge in ferrimagnets at the angular compensation point[18,19], synthetic antiferromagnets (SAF)[20–24], or antiferromagnets[25–31], where skyrmions can also achieve higher velocities than in ferromagnets.

Although racetrack memories[2,6,7] have been proposed as main application of magnetic skyrmions, their use for unconventional applications, such as in a reshuffle device for probabilistic computing[32,33], reservoir computing[34–36], true random number generators[37], Boolean logic computing gates[38–41], neuromorphic computing[42,43], artificial spiking neuron[44,45], and memristive networks[46–48] has shown very promising results.

In this work, we propose how to use magnetic skyrmions as a tool for the solution and visualization of the shortest path problem[49]. The main motivation is that many combinatorial optimization problems, which have application in real-life complex scenarios including transportation (e.g. autonomous vehicle navigation)[50], network optimization problems (e.g. dynamics programming, and internet routing)[51], and training of deep neural networks[52], require the evaluation of the shortest path. A number of mathematical algorithms[53] have been proposed to solve the problem of the shortest path, but the time for reaching the solution increases exponentially as the network becomes more complex. This calls for the development of different strategies to obtain a solution in a reasonable time. An approach to model such complicated system is to view the network as a number of interconnected resistors, which then correspond to a graph where the edges are resistors and the nodes are junctions between them. Each edge is characterized by a positive weight related to the resistance[54]. For instance, if we wish to study a transportation problem in terms of distance between cities, a high resistance indicates that two nodes (cities) are faraway, while a low resistance means that the nodes are close[54,55]. Alternatively, the resistors can be replaced by memristors, a passive dynamical two-terminal device introduced by Chua[56,57] and characterized by an input and history dependent variable resistance. Specifically, a larger current flow through a memristor leads to a larger change of the device resistance. Memristive networks[58] have been shown to solve more efficiently the problem of the shortest path by using the *analog computations* performed by solving Kirchhoff's Current Laws in a parallel manner[59]. However, to visualize the solution, one should know the potential for each node by including additional hardware to the system, such as voltmeters.

Here, we propose to take advantage of the magnetic skyrmions to visualize the solution of the shortest path in a 2-dimentional graph described as an electrical network, where the resistance of each resistor between two nodes is due to the length of the magnetic nanostrip (we consider a constant width here, but this can be used as additional degree of freedom to modulate the resistance values properly). Micromagnetic simulations are used for the theoretical proof of concept of this approach. We consider a graph made by a material stack which can host skyrmions, and two nodes of this graph $N_1$ and $N_2$. We apply a voltage between those nodes and the electrical current which gives rise to the SHE inside the HM flows in the graph with a spatial distribution which depends on its topology. Subsequently, skyrmions are periodically injected into $N_1$ and follow a path up to $N_2$. Therefore, the solution of the problem is achieved by using the final spatial distribution of the skyrmion train. If implemented in hardware, the computational time can be very efficient because it will depend on the skyrmion velocity which can reach Km/s[20,21]. Furthermore, the capability of skyrmions to move on interconnected grids along complex trajectories [1] is a key feature which makes them the only candidate over other magnetic solitons, e.g. domain walls and vortex, to be used in this category of optimization problems. Mainly, our analysis focuses on SAF 2D network because of the zero SHA, but we also show the negative effect of an oblique skyrmion trajectory by considering a single HM/FM heterostructure as a case of study. In addition, a comparison with Dijkstra's algorithm is included. Our predictions can be useful for the design of experimental systems exploiting skyrmions for the solution of the shortest path.

## 2. Methods: micromagnetic model

The micromagnetic study is performed by means of state-of-the-art home-made micromagnetic solver GPMagnet[60] which numerically integrates the Landau-Lifshitz-Gilbert-Slonczewski equation by applying the time solver scheme Adams-Bashforth, where the SHE is taken into account:

$$\frac{d\mathbf{m}}{d\tau} = -(\mathbf{m} \times \mathbf{h}_{eff}) + \alpha\left(\mathbf{m} \times \frac{d\mathbf{m}}{d\tau}\right) - \frac{g\mu_B \theta_{SH}}{2\gamma_0 e M_S^2 t_{FM}}\left[\mathbf{m} \times (\mathbf{m} \times (\hat{z} \times \mathbf{j}_{HM}))\right] \quad (1)$$

where $\mathbf{m} = \mathbf{M}/M_s$ is the normalized magnetization of the ferromagnet, and $\tau = \gamma_0 M_s t$ is the dimensionless time, with $\gamma_0$ being the gyromagnetic ratio, and $M_s$ the saturation magnetization. $\mathbf{h}_{eff}$ is the normalized effective field, which includes the exchange, IDMI, magnetostatic, and anisotropy fields $\alpha$ is the Gilbert damping, $g$ is the Landé factor, $\mu_B$ is the Bohr magneton, $\theta_{SH}$ is the spin-Hall angle, $e$ is the electron charge, $t_{FM}$ is the thickness of the ferromagnetic layer. $\hat{z}$ is the unit vector along the out-of-plane direction, and $\mathbf{j}_{HM}$ is the electrical current density flowing into the Pt heavy metal which gives rise to the SHE.

For the case of the SAF structure, two perpendicular FMs, separated by a thin Ruthenium (Ru), layer designed to provide an antiferromagnetic exchange coupling[61], are sandwiched between two different HMs. For this reason, in Eq. (1), $\mathbf{m}^i = \mathbf{M}^i/M_s^i$, where $i = L, U$ indicates the lower ferromagnet and upper ferromagnet, respectively. (see Ref. [21] for more details). The effective field includes the interlayer exchange coupling (same contribution for both FMs) $\mathbf{h}_{ex,i}^{inter} = \frac{A^{ex}}{\mu_0 M_s^{j2} t_{Ru}} \mathbf{m}^j$ where $i, j = L, U$, $A^{ex}$ is the interlayer exchange coupling constant, $\mu_0$ is the vacuum permeability, and $t_{Ru}$ is the thickness of the Ruthenium layer. The discretization cell used in the simulations is 4.0x4.0x0.8 nm$^3$. We consider typical parameters for a SAF structure[21,62]: $M_S$=600 kA/m, exchange constant $A$=10 pJ/m, $A^{ex}$ =-5.0x10$^4$ J/m$^2$, upper and lower IDMI parameters $D^L$=$D^U$=1.8 mJ/m$^2$, perpendicular anisotropy constant $k_u$=0.50 MJ/m$^3$, $\alpha$ =0.1, and $\theta_{SH}^L = 0.1$ and $\theta_{SH}^U = 0$.

## 3. Results

*3.1. Néel Skyrmions in SAF*

We analyze the skyrmion motion in a 2-dimensional system (Fig. 1) characterized by an area occupancy of L$^2$, with L=2700 nm being the length of the system, where each nanowire has a width w=100 nm. The two HMs have a thickness $t_{HM}$=3 nm, while $t_{FM}$ and $t_{Ru}$ are 0.8 nm.

We computed the spatial distribution of the current[63,64] (see Fig. A1) by supposing to supply voltage between the terminals N$_1$ and N$_2$ (Fig. 1(a)). The spin-current is polarized perpendicularly to the electrical current. We continuously nucleate a single skyrmion in correspondence of the node N$_1$, then the spin-orbit-torque originating from the SHE drives the skyrmions towards the node N$_2$. For the SAF structure, we consider only one current flowing through the lower HM[21]. We inject a current |$j_{HM}$|=3 x 10$^8$ A/cm$^2$ [21], and nucleate a skyrmion every 10

ns. This nucleation time has been properly designed in order to avoid skyrmion-skyrmion interactions during the motion. In other words, each skyrmion is well separated by the previous and next ones. The results are robust on a wide range of current $2 \times 10^8$ A/cm$^2$ < $|j_{HM}|$ < $4 \times 10^8$ A/cm$^2$ with a proper choice of the skyrmion nucleation time.

The nucleated skyrmion moves by following the current distribution. If it meets an intersection where the current splits equally, the path selection is deterministic and the skyrmion tries to keep the same trajectory owed to an inertia effect deriving from the motion along the first part of the track (see Movie 1). For this specific device, after about 130 ns, the solution emerges, as we can see in Fig. 2. This can be read by using state-of-the-art imaging techniques [18,22]. Basically, each new skyrmion introduced into the system will go through the same path analogously with ant colony algorithm where the ant agents follow the path with a larger amount of pheromone[65–67]. The fundamental aspect to be underlined is that we do not need to know locally the value of the current and/or of the voltage, but it is the overall current distribution, which is linked to the geometry of the system, to allow the solution to emerge and be detected by simple imaging techniques. This is the main achievement of this work. Our results can be extended to skyrmion in antiferromagnets[25–28] and compensated ferrimagnets[18,19] because they are characterized by a zero SHA.

*3.2. Skyrmions in ferromagnets*

We stress one more time that the SAF structure is a suitable solution to envisage applications involving skyrmion dynamics due to the zero SHA[20,21]. The need of a zero SHA is immediately clear when extending this study to single HM/FM bilayer where Néel skyrmions are stabilized by the IDMI due to the HM/FM interface[1,2,68]. The SHE-driven Néel skyrmion motion is well-known[7] and it is characterized by a finite SHA $\phi_{SkH} = arctg \frac{v_y}{v_x} = arctg \frac{S}{\alpha \mathcal{D}}$, with $v_x$ and $v_y$ being the x- and y-components of the skyrmion velocity, $S = \frac{1}{4\pi} \int \mathbf{m} \cdot (\partial_x \mathbf{m} \times \partial_y \mathbf{m}) dx dy$, $\alpha$ is the Gilbert damping, and $\mathcal{D}$ is the dissipative tensor [7,16]. We consider the same geometry as in Fig. 1 and typical parameters of single HM/FM bilayer (e.g. Pt/CoFeB)[7] $M_S$=1000 kA/m, $A$=10 pJ/m, $D$=2.0 mJ/m$^2$, $k_u$=0.80 MJ/m$^3$, $\alpha$ =0.1, and $\theta_{SH} = 0.1$. We nucleate Néel skyrmions in correspondence of the node N$_1$ and we move them by means of the SHE. The value of the current amplitude is $|j_{HM}|$=5 MA/cm$^2$, while the nucleation time is 10 ns to obtain well-separated skyrmions. After being nucleated, the skyrmion moves along the positive x and y directions and, differently from the SAF case, here when the skyrmion meets the first intersection, the selection of the path is due to both the SHA and the current distribution. In fact, the skyrmion does not follow the path with larger current, as in the SAF case, but it goes through the vertical path because of the large SHA (see Movie 2). When the skyrmion goes to the second intersection, its motion is characterized by a positive y-component and a negative x-component of the velocity. Hence, at the second intersection, it stops because of the absence of the current (see Movie 2). In this way, an accumulation of steady skyrmions forms in the part of the track where there is no current. The next skyrmion going towards the second intersection is then repulsed by the magnetostatic interaction arisen from the steady skyrmions and it keeps going up. A similar behavior will then occur at each intersection. However, this mechanism involving a finite SHA and skyrmion-skyrmion interactions does not allow skyrmions to follow the shortest path, as shown in the final configuration of Fig. 3. We hence conclude that the skyrmion motion characterized by a finite SHA fails to solve those problems.

*3.3. Skyrmions in other systems*

The previous results demonstrate that only systems where the SHA is suppressed can be used for the solution of the of the shortest path problem with skyrmions. In addition to SAFs[20,21], antiferromagnets[25–28] and compensated ferrimagnets[18,19], a further viable scenario concerns the use of STT-driven skyrmions[7] in a single ferromagnet. Such a motion is still characterized by a finite SHA, but this can be controlled by the $\frac{\beta}{\alpha}$ ratio[1,6,7,12], where $\beta$ is the non-adiabatic parameter. In particular, if $\beta$ is properly designed to be equal to $\alpha$, the skyrmion motion exhibits a zero SHA[1,6,7,12]. Therefore, we would obtain a final solution similar to the one in a SAF (Fig. 2), where the skyrmions would follow the current distribution. In this case, the achievement of the solution would be slower than SAF case due to the slower STT-driven skyrmions motion, while the imaging of the finale state will be easier.

Recently, a growing interest have been devoted to magnetic multilayers[69–74], where tiny "hybrid skyrmions" (size smaller than 100 nm) can be stabilized at room temperature in absence of IDMI. Such hybrid skyrmions are characterized by a thickness-dependent reorientation of their domain wall chirality and the equilibrium configuration depends on the interplay between dipolar interactions and IDMI. In particular, if the dipolar interactions are the dominant contribution, a Néel skyrmion with outward chirality in the top ferromagnetic layer will be gradually reoriented to a Bloch skyrmion in the middle ferromagnetic layer and to a Néel skyrmion with inward chirality in the bottom ferromagnetic layer[72,73]. On the other hand, if the IDMI is increased, the Bloch skyrmion is shifted towards the upper layers, until, beyond a threshold IDMI value, a pure Néel skyrmion is obtained along the whole thickness[72,73]. Theoretical studies have pointed out that the SHE-driven skyrmion motion in these multilayers is characterized by a SHA dependent on the value of the DMI and/or on the number of repetitions of the ferromagnetic layer[72,75]. Therefore, by proper designing the IDMI and/or the number of repetitions, it is possible to achieve a zero or at least negligible SHA, thus making multilayers a desirable system for the solution of the of the shortest path problem with skyrmions. If, on one hand, the realization of skyrmion-hosting multilayers is well-established[69–71,73], on the other hand, skyrmions undergo a strong pinning effect while driven by the SHE[69]. Besides the use of skyrmions in different systems, it would also possible to use a different magnetic soliton, such as antiskyrmion [76,77] or skyrmionium[78,79], as long as the SHA is zero.

*3.4. Comparison with the Dijkstra's Label Algorithm*

To conclude our work, Fig. 4 shows a comparison of the proposed algorithm based on current distribution and skyrmions (green line) with the Matlab® implementation of the Shortest Path Dijkstra's Label Algorithm (red line) considering a graph with undirected edges and nodes having the same topology as the system of Fig. 1. The edge weights are computed by considering the resistance of the physical device for each track, which is a function of the material, width and length of each track. A related dimensionless parameter is applied along each edge to reference its corresponding weight. The node 17 is the source node, which corresponds to node $N_1$ in Fig. 1, and the node 84 is target node ($N_2$ in Fig. 1).

It can be observed that the two solutions do not coincide but both are optimal solutions if one considers the sum of the total length of the graph. If we increase the resistance value between node 17 and 18 of 1, the Shortest Path Dijkstra's Label Algorithm also converges in the green line (not shown). Therefore, our benchmark against the Dijkstra's Algorithm confirms the validity of our skyrmion-based approach.

## 4. Perspectives and conclusions

The results achieved here have been obtained by means of a micromagnetic model taking into account only the current distribution (purely resistive behavior) which does not depend on the magnetic texture, i.e. on the presence/absence of skyrmions. Indeed, experimental measurements have shown that the resistance of a ferromagnetic material depends on the number of skyrmions[71]. This outcome is very promising for our aim to solve the shortest path problem with skyrmions because a memristive behavior can occur. Specifically, when a skyrmion starts to follow a certain path, the electrical resistance of that path decreases, leading to an increase of the current. This aspect induces the next skyrmion introduced into the system to trail the previous skyrmion, and thus, to further reduce the electrical resistance of the selected path. Eventually, a self-reinforced path emerges, as it happens for real ants seeking for food as well as in the ant colony optimization algorithm[67].

In summary, we have shown, by means of micromagnetic simulations, that skyrmions can be used to display the final solution of the shortest path problem by exploiting the current distribution in a generic 2D graph. The key ingredient is to consider systems where the SHA is zero or negligible, such SAFs, while the solution cannot be achieved if the SHA is finite, such as in single ferromagnets. Our proposal is even more appealing if experimentally implemented due to the decreasing of the electrical resistance with the number of skyrmions. The combination of our theoretical and experimental results opens the way for the hardware realization of skyrmion devices for the solution of the shortest path problem based on a self-reinforced algorithm.

## Acknowledgements

This work was supported by the project "ThunderSKY", funded by the Hellenic Foundation for Research and Innovation (HFRI) and the General Secretariat for Research and Technology (GSRT), under grant agreement No 871.

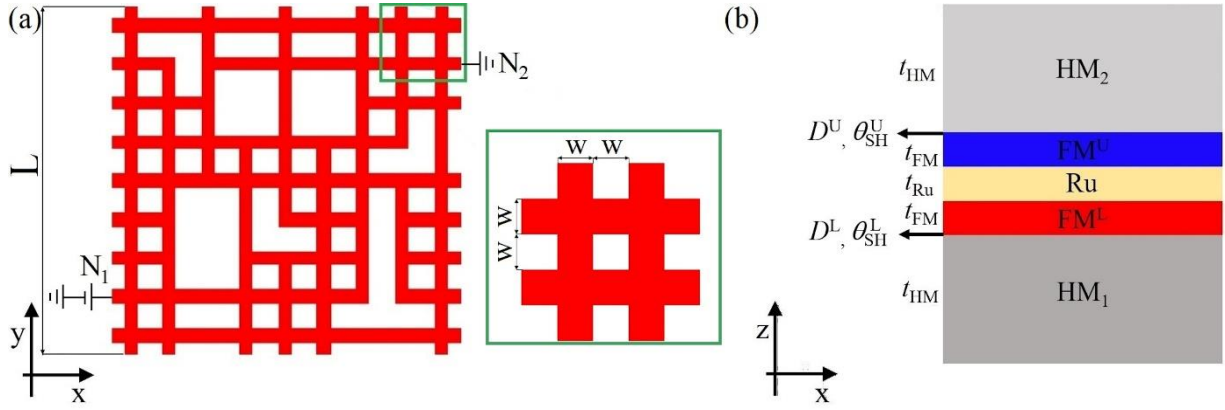

**Fig. 1**. Sketch of the device under investigation where (a) shows the x-y view together with a magnification highlighted by a green square and indicating the width of the structure and (b) displays the cross-section along the z-direction.

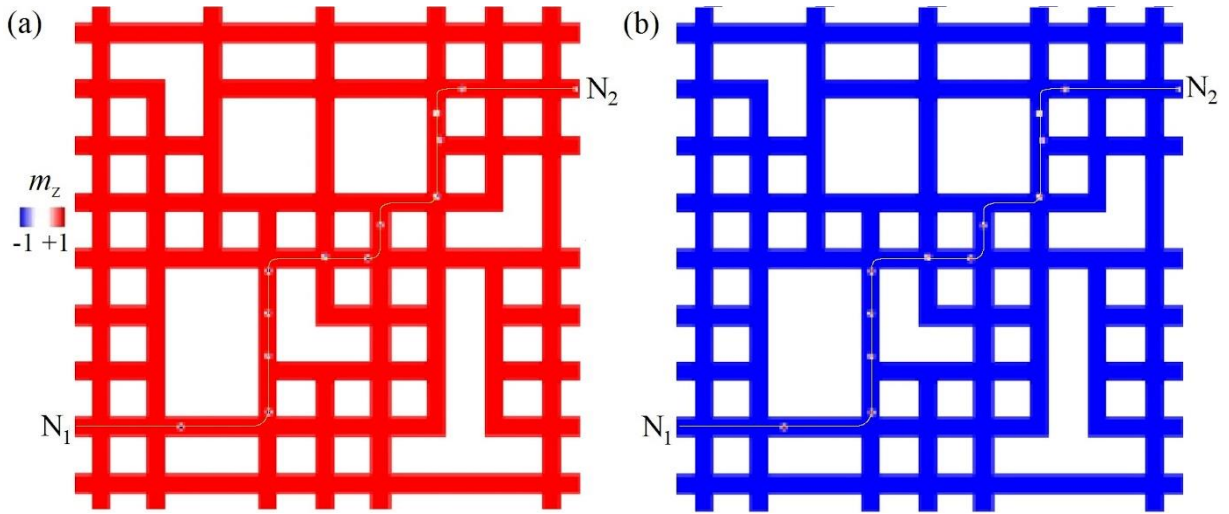

**Fig. 2.** Spatial distribution of the magnetization for the (a) lower and (b) upper layer of the SAF structure[21,22]. The path identified by the skyrmion train and indicated with a yellow solid line represents the final solution of the problem.

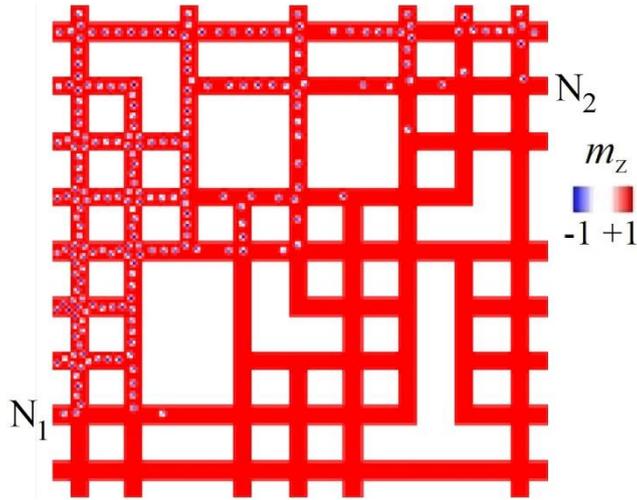

**Fig. 3.** Final spatial distribution of the magnetization for the skyrmions in a single HM/FM bilayer.

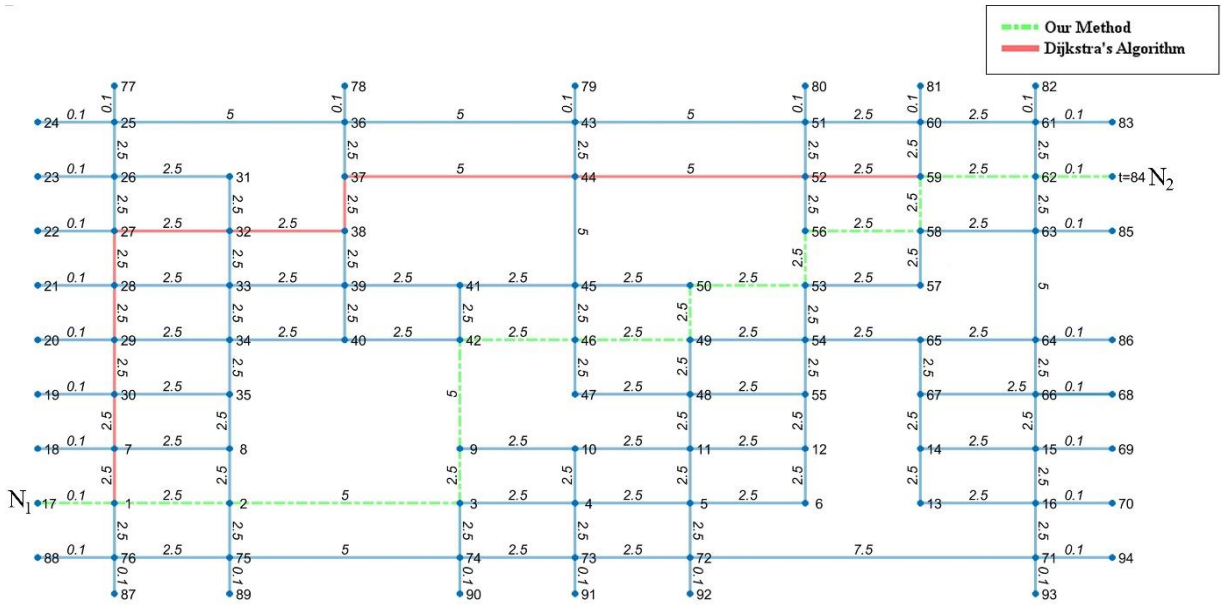

**Fig. 4**. Graph with undirected edges as obtained by modeling the same structure as Fig. 1. Nodes and edge locations and distances are accurately reproduced. Edge weights are represented by the linear resistance of the corresponding track. The path as obtained using our algorithm (dashed green line) closely resembles the one obtained from numerical simulations, whereas the path obtained using Dijkstra's Algorithm (solid red line) is different.

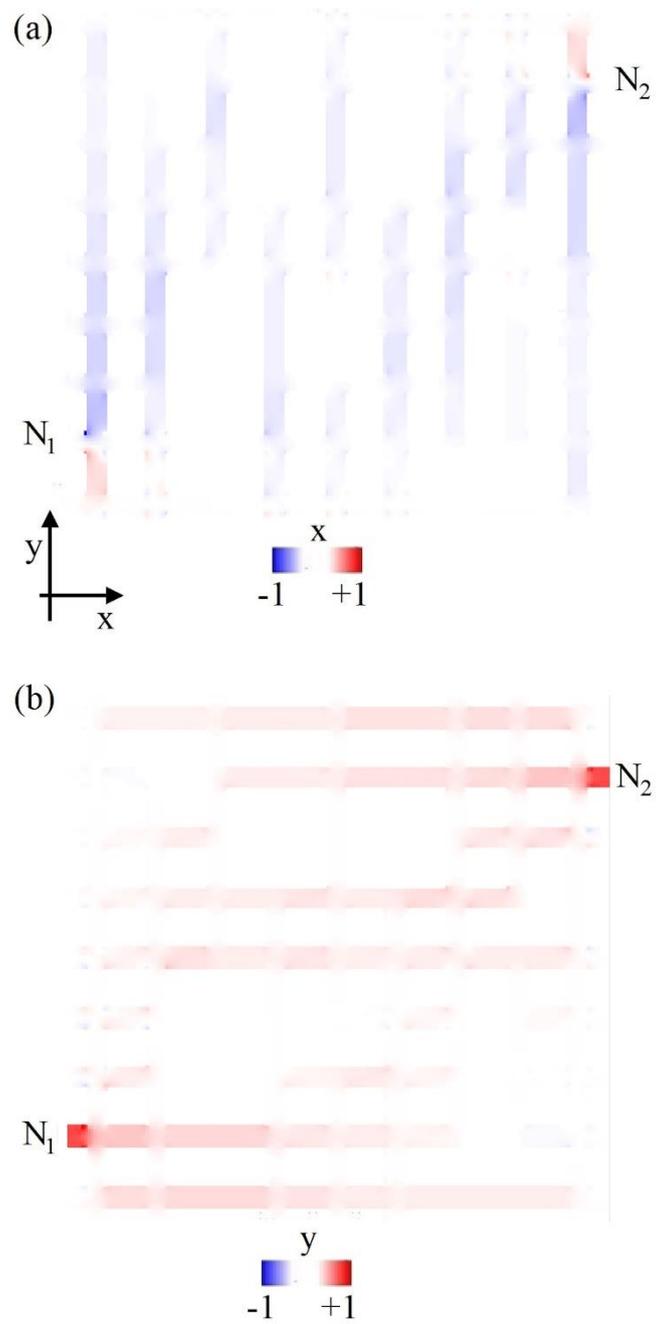

**Fig. A1**. Spin-current spatial distribution where (a) and (b) show the x- and y-components, respectively. Where the spin-current is along the x- (y-) direction, the skyrmions move vertically (horizontally).